\documentclass[acmsmall,nonacm]{acmart}

\setcopyright{acmcopyright}
\copyrightyear{2023}
\acmYear{2023}
\acmDOI{XXXXXXX.XXXXXXX}

\acmConference[SIGMOD]{Make sure to enter the correct
	conference title from your rights confirmation email}{
	2024}{}
\acmPrice{15.00}
\acmISBN{978-1-4503-XXXX-X/18/06}

\newtheorem{example}{Example}%[section]
\usepackage{subcaption}

\newif\ifarxiv
\arxivfalse

\begin{document}
	
	\title{DTT: An Example-Driven Tabular Transformer for Joinability by Leveraging Large Language Models}
	
	\author{Arash Dargahi Nobari}
	\email{dargahi@ualberta.ca}
	\affiliation{%
		\institution{University of Alberta}
		\city{Edmonton}
		\state{Alberta}
		\country{Canada}
	}
	
	\author{Davood Rafiei}
	\email{drafiei@ualberta.ca}
	\affiliation{%
		\institution{University of Alberta}
		\city{Edmonton}
		\state{Alberta}
		\country{Canada}
	}

	\begin{abstract}
		Many organizations rely on data from government and third-party sources, and those sources rarely follow the same data formatting. This introduces challenges in integrating data from multiple sources or aligning external sources with internal databases. 
		Commercial database systems do not offer adequate support for integrating data from heterogeneous sources, and manual integration is both time-consuming and inefficient. State-of-the-art data integration approaches that rely on similarity functions and textual transformations often fail to handle challenging cases where multiple mappings are required, or the mappings go beyond simple textual transformations.
		
		In this paper, we study the potentials of deep neural models for transforming tables for joinability. In particular, we cast the problem as a prediction task and develop a framework that leverages large deep-learning language models to transform tabular data from a source formatting to a desired target representation. Our framework can efficiently learn the patterns for mapping a source formatting into an expected target using just a few examples, which can then be used for tasks such as table joining, filling in missing values, and error detection.
		Compared to state-of-the-art mapping and joining approaches, our framework delivers noticeably more accurate and scalable performance on both real-world and synthetic datasets. Our experimental evaluation also shows that the performance of the proposed framework using our fine-tuned model is at par or better than large language models such as GPT-3, despite the significant difference in size, and that using large language models within our framework improves their performance.
	\end{abstract}
	
	\maketitle

	\section{Introduction}
	The drive towards data publishing and sharing by entities and governments over the past couple of years has led many organizations to rely on data from third-party sources. However, gathering data from multiple sources inevitably leads to data mismatches. Converting data from one format to another has long been a challenge in data integration and management~\cite{FlashFill,FlashFill2,autotransform,TDE}, with traditional approaches relying on manual development of guidelines and rules for integration or transformation~\cite{ETL:2005,data:2003}. However, the sheer size and complexity of modern databases make manual transformation and integration impractical, fueling a significant research on automatic data transformation~\cite{icde,autojoin,tde_demo,foofah}. 
	
	\begin{figure}[tbp]
		\centering
		\includegraphics[width=.55\linewidth]{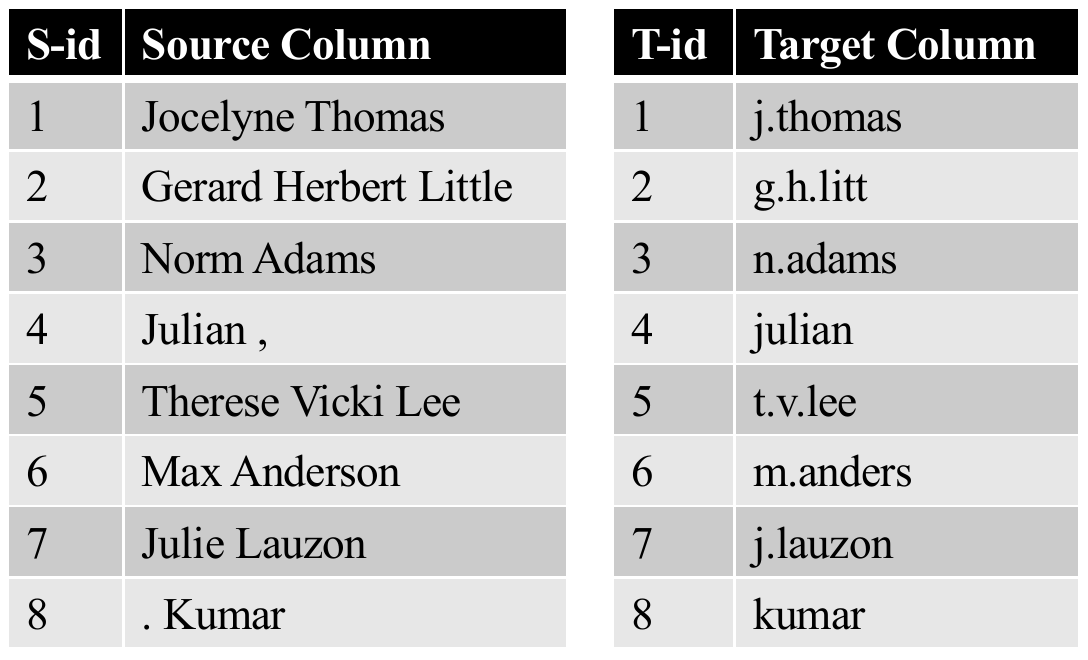}
		\caption{Two columns, representing the name and user id of individuals, where a mapping from name to user id is sought}
		\label{fig:tables}
	\end{figure}

	Our focus in this paper is on automated transformation of tabular data (e.g.,  spreadsheets, web tables, and relational databases), which are widely adopted by both organizations and governments for representing, storing, and sharing data.
	In particular, given a few examples of matched rows between a source and a target, the goal is to learn a mapping. Those mappings can then be used to transform arbitrary rows in the source formatting into the formatting of the target, with applications in joining data from different sources~\cite{icde,autojoin}, filling in the missing values and auto-completion~\cite{BlinkFill,FlashFill}, and error detection and correction~\cite{he2016interactive}.

	\begin{example}
		\label{eg:eg1}
		Figure~\ref{fig:tables} depicts a source and a target, representing the same entities (people) but in different formattings. The source table shows the names of the individuals, while the target table shows their corresponding user ids. Suppose that the target column is incomplete or unavailable, and the objective is to predict the missing values of the target column, based on a few examples of source-target pairs. Or, one may want to join the two columns despite the differences in formatting. The transformation process requires some reformatting rules and the choice of a rule may be conditional on the input. For example, in the first, third, and seventh rows, the reformatting rule involves concatenating the initial and the last name, converting them to lowercase, and using a period as a separator. However, in the second row, there is a middle name, while the last row lacks a first name, and these variations can affect the transformations.
		In general, detecting such transformations from a set of examples is not straightforward and must account for various challenges.
	\end{example}
	
	\noindent
	\textbf{Challenges}
	The \textit{search space} for possible transformations is huge. If each transformation is synthesized into a sequence of edit operations, the search space grows exponentially with the number of edit operations in a transformation and the parameter space of the operations. Also, both the search space and the runtime for many state-of-the-art approaches~\cite{icde,autojoin,BlinkFill,FlashFill} further grow with the input size, such as the number of rows and the length of a row. Despite some attempts to reduce the search space by limiting the number of operations (e.g., split and substring) within a transformation~\cite{icde}, sampling the input~\cite{autojoin}, and applying pruning strategies~\cite{icde}, the search space still remains relatively large. Hence the time needed to find a mapping is often much more than what may be considered as acceptable, for example, in an online setting. 
	Also, some of these improvements and prunings are lossy, and they can miss transformations that are of a better quality than those found.

	Another challenge is \textit{the availability of input examples and noise handling}.
	The examples are usually either user-provided or automatically generated from input. In the former, the examples are (extremely) limited but accurate, whereas in the latter, the examples can be extensive, but less accurate. In real-world settings, noise is usually unavoidable and inconsistencies can exist in data. Also when examples are automatically generated, some of them can be incorrect. 
	A good model should perform well with only a limited number of examples, and it should also benefit from the abundance of examples, maybe ignoring those that are less useful.
	A good model should also be robust against any possible noise in data and deal with inaccuracy in the provided examples.
	
	\noindent\textbf{Existing Approaches}
	A wide array of studies target the problem of matching entity records that describe the same real-world entities but differ in terms of formatting or representation~\cite{auto-em,auto-fuzzy-join,Li2020:Deep,MassJoin,semajoin,wang2017synthesizing,akbarian2022probing}. Traditional approaches rely on textual and semantic similarity, whereas more recent approaches incorporate machine learning and deep neural models. While these models provide effective solutions to the problem of formatting mismatch in joining tabular data, their use cases are limited. For example, these models cannot predict missing values, provide suggestions, or detect outliers in a table. This has led to another line of research where the focus is on finding a mapping between source and target tables and leveraging this mapping to transform the source formatting into that of the target.
	
	The majority of approaches aimed at detecting a mapping between two entity columns heavily rely on a limited set of string-based transformations~\cite{icde,autojoin,BlinkFill,FlashFill} and an exhaustive search of the parameter space. 
	While the search space can be bounded by limiting the number of string-based transformations, this can negatively impact the accuracy.
	Consider the source and target tables in Figure~\ref{fig:tables}, where different rows may require different formatting rules. To transform all rows in the source to their corresponding target values, six distinct textual transformations may be needed, as illustrated in Example~\ref{eg:eg1}. Some studies~\cite{autojoin,FlashFill,BlinkFill} limit their search space and find a single transformation that covers all input rows, which will not be effective in this scenario. Other studies~\cite{icde,TDE} can produce more than one transformation, but the problem of selecting a transformation from the set to apply to an arbitrary row is left unanswered. For instance, Nobari~et~al.~\cite{icde} provide a set of transformations that are required for a mapping but do not provide much hint on how to select a transformation for an input row.
	Furthermore, many state-of-the-art methods~\cite{icde,autojoin,BlinkFill,FlashFill} exhaustively search the transformation space, and despite their pruning strategies, their runtimes increase dramatically when the input size grows~\cite{icde}.
	
	\noindent\textbf{Our Approach}
	In this paper, we introduce Deep Tabular Transformer (DTT), a novel framework for transforming tabular data into a joinable format using the power of deep learning for language modeling. Similar to the approaches in the literature, DTT focuses on transforming entity columns. However, unlike traditional approaches that are limited to a few pre-defined string-based transformations and an exhaustive search process, DTT overcomes these limitations by leveraging advanced deep learning techniques. DTT predicts an expected output row in the target table for each row of the source table, enabling easy and efficient data joining. While the focus of this study is the utilization of DTT for single column joining, the framework can be extended and adapted for various other downstream tasks.
	Our experimental results show that DTT outperforms existing state-of-the-art approaches in terms of accuracy, is applicable to a larger set of tables, and maintains an outstanding runtime performance even when dealing with large input size.  Remarkably, the performance of DTT is at par or better than large language models such as GPT-3 despite having an order of magnitude less parameters and requiring dramatically less resources during inference. We are releasing DTT as a pretrained model, which demonstrates exceptional performance across multiple domains without the need for fine-tuning. Our hope is that this release will drive further advancements in the field.

	Our contributions can be summarized as follows:
	\begin{enumerate}
		\item We propose DTT, a novel example-driven approach for the transformation of tabular data, leveraging the capabilities of pretrained language models.
		
		\item We develop a diverse dataset for training our model, comprising of synthetic examples. Our experiments demonstrate that our model performs exceptionally well on real-world data from various domains.
		
		\item We present an end-to-end framework for tabular data transformation which includes a decomposer, serializer, model, and aggregator. As an application of our framework, we demonstrate its effectiveness in table joining.
		
		\item Through an extensive evaluation conducted on a wide range of datasets from different domains, we show that our approach outperforms existing state-of-the-art baselines in terms of both accuracy and runtime.
		
		\item We make all our resources, including our code, framework, pretrained model, synthetic data generator, and real-world benchmarks publicly available for the research community\footnote{https://github.com/arashdn/dtt/}.
		
	\end{enumerate}

	\section{Problem definition}
	\label{sec:problem_def}
	We want to transform tables from a source formatting to a target formatting using a few provided examples.
	Let $S=\{s_1, s_2, \ldots \}$ denote a set of values in the source. For a small subset $S^\prime \subset S$, let $E=\{(s_i,t_i) | s_i \in S^\prime\}$ denote a set of $k$ examples where the target values are given to guide the process for finding a transformation. The aim is to find the target formatting of every value in the source, i.e.
	\begin{equation}
		\label{eq:target}
		R = \{ (s_i, f(s_i)) | s_i \in S \wedge \forall s_j \in S^\prime ((s_j, f(s_j) \in E)\}.
	\end{equation}

	As an example, suppose we have a source table $S$ that lists the recent prime ministers of Canada and an example set $E$ that consists of three rows:
	\\
	\texttt{ 
		S = \{`Justin Trudeau', `Stephen Harper', `Paul Martin', \\
		`Jean Chretien', `Kim Campbell'\},\\
		E = \{
		(`Justin Trudeau', `jtrudeau'),\\ 
		$\text{}$\hspace{27px}(`Stephen Harper', `sharper'),\\
		$\text{}$\hspace{27px}(`Paul Martin', `pmartin')
		\}.
	}\\
	Our aim is to find the target formatting  for any arbitrary value in $S$. For instance, the values \texttt{`Jean Chretien'} and \texttt{`Kim Campbell'} may be mapped to \texttt{`jchretien'} and \texttt{`kcampbell'} respectively.

	Tables can be transformed for joinability, for example, allowing a column of a source table to be joined with a column in the target. Tables may also be transformed to fill in missing values in a target column. In both cases, $S$ can be the set of all values in the source column.
	In this study, we assume the source values and examples are provided. This is a common practice to limit the scope of the problem and focus on data transformations~\cite{icde}. If user-provided examples are not available, an unequal joining method~\cite{auto-fuzzy-join,MFJoin,semajoin,wang2017synthesizing} or token-based example generation~\cite{icde,autojoin} may be used to provide a set of examples, with the caveat that the automatically generated examples may contain noise and invalid pairs. We will discuss how our approach can deal with such noisy examples.

	\section{Background and Related Work}
	Our work is related to the lines of work on (1) example-driven tabular data transformation, (2) language modeling and text-to-text transformers, and (3) language models applied to tabular data. This section reviews the relevant works in these areas, while filling the gaps with some background (when necessary).
	
	\subsection{Example-Driven Tabular Data Transformation} 
	This is probably the closest line of work to ours.
	There are numerous studies in this area~\cite{icde,autojoin,BlinkFill,FlashFill,FlashFill2}, and FlashFill~\cite{FlashFill} and BlinkFill~\cite{BlinkFill} are among the pioneers, with a focus on spreadsheet data. These two approaches construct an input graph, based on a given set of user-provided examples, which is then traversed to generate a sequence of substring-based textual transformations that map source values to their corresponding targets. However, FlashFill and BlinkFill heavily rely on the accuracy of the provided examples and are unable to handle noise in the examples. To address this issue, Zhu~et~al. propose a method called Auto-join~\cite{autojoin}, which uses a set of pre-defined string-based transformation units, such as substring and split, to describe the transformations. The examples are automatically generated by token matching, and the method creates several subsets of the input examples to handle noise. A recursive backtracking algorithm is then applied to each subset to find the best transformation. While Auto-join is able to handle minor noise in the input and limits the search space by using pre-defined transformation units, it is a backtracking method and needs to search the entire transformation space in the worst case, which can be computationally expensive. Also, it may not perform well if the noise level in the examples is significant. 
	
	In a more recent work by Nobari~et~al.~\cite{icde}, referred to as Common String-based Transformer (CST), the search space for string-based transformations is further constrained by considering common text sequences between source and target examples as textual evidence to form the skeleton of transformations. CST uses the same string-based transformation units as Auto-join, but transformations for each row are generated independently to better handle input noise. The transformations are then ranked based on their coverage to build a final transformation set. CST offers better noise handling and runtime performance compared to Auto-join, but it is still limited to substring-based transformation units and performs well only when long matching sequences exist between source and target examples. 
	Although the pruning conditions in Auto-join and CST limit their search space and improve their runtime, they can end up missing some transformations, particularly those that cannot be covered by a small set of pre-defined transformation units. Our aim is to overcome these limitations on the search space by utilizing a Language Model (LM) to transform source values into the desired target representation.

	\subsection{Language Modeling and Text to Text Transformers}
	With large language models forming an integral component of our framework, we provide a brief background of those models.
	A vast majority of machine-learned LMs are based on the concept of masked language modeling, where some tokens in a given sequence are masked (or corrupted), and the model is trained to predict those masked tokens.
	Word2Vec~\cite{word2vec} and GloVe~\cite{glove} are among the earliest models for pretraining, which generate static vectorized embeddings of each word using a shallow neural network. A later work, ELMo~\cite{elmo}, uses two layers of bidirectional LSTM~\cite{lstm} to observe the context before and after a word and generates contextualized embeddings of the words, unlike the static embeddings in Word2Vec. In recent years, Vaswani~et~al.~\cite{transformers} introduce transformers that use self~attention~\cite{transformers}, allowing the model to parallelize better than LSTM models and not giving more weight to nearby words. Transformer-based models consist mostly of an encoder, a decoder~\cite{seq2seq_NIPS2014}, or both. Encoder-only models, such as BERT~\cite{bert}, aim to learn the natural languages and generate a latent representation of the input that can be used for tasks requiring an understanding of the input. Decoder-only models, such as GPT-2~\cite{gpt2} and GPT-3~\cite{gpt3}, are widely used to generate natural language text given a context. Finally, Encoder-Decoder models, also referred to as sequence-to-sequence or text-to-text models, such as T5~\cite{t5} and BART~\cite{bart}, use an encoder to create a latent representation of the input, which is passed to the decoder to generate a new text for a desired task.
	
	\subsection{Language Models Applied to Tabular Data}
	The proliferation of pretrained language models has resulted in a wide array of transformer models being applied to table embedding and comprehension~\cite{tacl_ttdr, dong2022table, katsogiannis2023survey}, alongside their growing utilization in diverse tasks involving tabular data.
	Several of these models, including TaBERT~\cite{tabert}, TURL~\cite{turl}, and TABBIE~\cite{tabbie}, employ an encoder-only architecture, rendering them more suitable for discriminative tasks such as entity matching and question answering. On the other hand, models such as RPT~\cite{rpt} employ an encoder-decoder structure, making them preferable for generative tasks. Our proposed model also follows an encoder-decoder architecture, specifically tailored for the generation task in DTT, as further discussed in Section~\ref{sec:model}.
	In terms of use cases, these models are being applied to different data processing tasks~\cite{quamar2022natural,LRR23} such as entity matching~\cite{DITTO,akbarian2022probing},	text to SQL~\cite{wang-etal-2020-rat,katsogiannis2023survey},
	question answering~\cite{chen2020open,chen2022large,tabert,tabbie} and 
	data to text~\cite{parikh-etal-2020-totto,kale-rastogi-2020-text,thorne2021natural}. 
	%Many deep learning and NLP models can only process data as a sequence of tokens, hence 
	These tasks are orthogonal to ours with table serialization becoming a common module in many of these tasks.
	Several serialization techniques have been developed to transform tables into sequences of tokens~\cite{tabert,turl,rpt,tabbie}, while preserving the structural relationships that may be needed for these tasks. Since the relationships that need to be preserved can be task-dependent, the choice of a  serialization method can also be task-dependent. For example, Iida~et~al.~\cite{tabbie} pass the rows and columns as two separate sequences to two transformer blocks and average the row and column values for each cell to generate a cell representation. In RPT~\cite{rpt}, tables are serialized using two special tokens, \texttt{[A]} and \texttt{[V]}, to encode attribute names and their corresponding values respectively. While this serialization keeps the structural information about the tables, it is not very efficient as the attribute names are repeated in each row of the table. Our aim is not to generate a dense representation of the entire table, and this requires a different serialization approach, which we discuss in Section~\ref{sec:serialization}.

	\section{approach}
\begin{figure*}[tbp]
		\centering
		\includegraphics[width=1\linewidth]{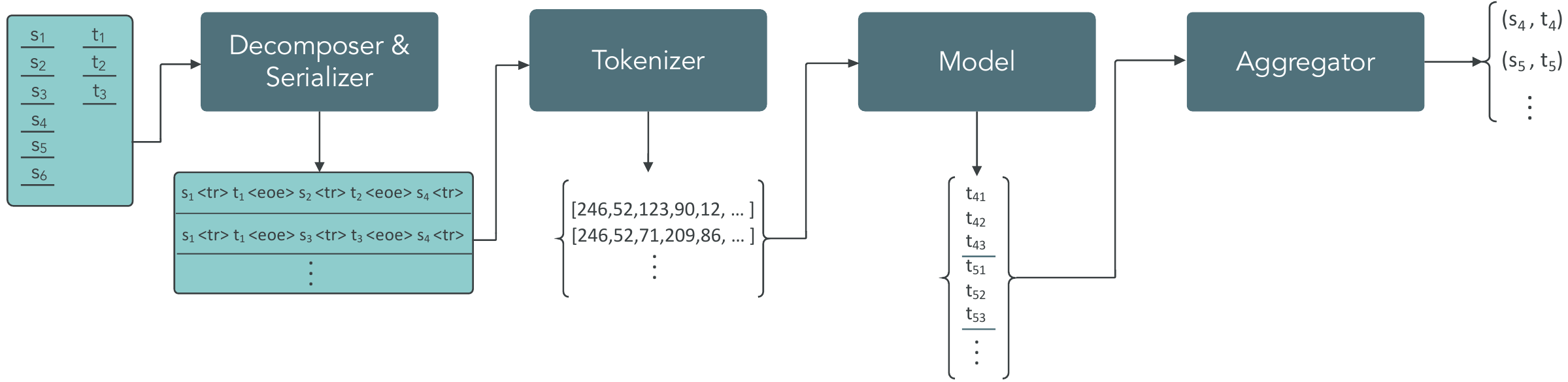}
		\caption{The architecture of DTT}
		\label{fig:framework}
	\end{figure*}

	As depicted in Figure~\ref{fig:framework}, our framework consists of a few components: (1) a decomposer and serializer, which decomposes the problem into smaller subtasks and performs an input serialization; (2) a tokenizer, which performs tokenization to obtain a vectorized representation of the input; (3) a sequence-to-sequence model, which predicts an output for each subtask; and (4) an aggregator, which is responsible for combining the predictions of the subtasks to generate a  final prediction.
	In the rest of this section, we will discuss the details of those components.
	
	\subsection{Decomposer and Serializer}
	\label{sec:serialization}
	
	Given a set of rows $S$ from a source table and an example set $E$ of source-target row pairs, the aim is to transform every row $s_i \in S$ to a target based on the examples. Many Large Language Models (LLM) impose a limit on the length of the input. This limit usually varies from 512 tokens (e.g., BERT) to 2048 tokens (e.g., GPT-3) and is associated with the quadratic memory requirement of the self-attention mechanism with input length~\cite{zaheer2020big}. However, the encoding of a table can be much longer for large tables and when there are many examples. 
	%Furthermore, it has been shown that the output generated by LLMs may not be consistent given a single prompt~\cite{wang2022self,yao2023tree}. Wang~et~al.~\cite{wang2022self} suggest utilizing LLMs with more than one prompt and applying a majority voting to obtain final results, referred to as self-consistency.
	To reduce this dependency on input length, we decompose the problem into smaller tasks, with each task being small enough to easily fit the input length requirement of many language models. This process enables the framework to prompt the model multiple times, each with different context samples, and this can further improve the model self-consistency. It has been shown in other domains that the output generated by LLMs may not be consistent given a single prompt and that multiple prompting is expected to help~\cite{wang2022self,yao2023tree}. One may also prompt multiple models instead of a single one, for example a model fined-tuned for the task and a general purpose LLM, to further improve the consistency and accuracy of the outputs (see Section~\ref{sec-multi-model}).
	This decomposition process is discussed in this section and the aggregation of the results is discussed in Section~\ref{sec:aggregator}.

	Suppose the number of examples that describe the context in a sub-problem is set to two. For any arbitrary row in the source table, any subset of $E$ of size two can be selected as the context. Let $E^2$ denote the set of all subsets of $E$
	of size two, i.e.
	\begin{equation}
		E^2 = \{(s_1, o_1), (s_2, o_2) | (s_1, o_1) \in E \wedge (s_2, o_2) \in E \wedge s_1 < s_2\}.
	\end{equation}

	For each input row $s_i \in S$ to be transformed, there are $|E^2|$ possible contexts that can be chosen.
	As an example, consider sets $S$ and $E$ in Section~\ref{sec:problem_def}. The set $E^2$ of all subsets of size two of $E$ will be
	\texttt{ \small
		E$^2$ = \{\\
		<(`Justin Trudeau',`jtrudeau'),(`Stephen Harper',`sharper')>,\\
		<(`Justin Trudeau',`jtrudeau'),(`Paul Martin',`pmartin')>,\\
		<(`Paul Martin',`pmartin'),(`Stephen Harper',`sharper')>\},\\
	}
	and an encoding of the input \texttt{`Jean Chretien'} $\in S$ using one of these contexts is 
	\texttt{ <(`Justin Trudeau', `jtrudeau'), (`Paul Martin', `pmartin'), (`Jean Chretien',>}.
	Each input row $s_i$ can be fed to the model multiple times, each time with a different context. If the input is passed to the model $n$ times, each time with a different context, the model will predict $n$ possible targets, which can be aggregated as discussed in Section~\ref{sec:aggregator}.
	
	It is common to use special tokens to mark the beginning and the end of sentences in natural language as it helps with training a model. The same convention is followed in encoding tabular data to describe the relationships between different input fields.
	Following this convention, we separate the source and target in an example with a \texttt{<tr>} token and two examples with \texttt{<eoe>}. We also mark the beginning of input with \texttt{<sos>} and the end of input with \texttt{<eos>}.
	With these symbols, our example given earlier can be encoded as:
	\texttt{
		\small
		<sos>Justin Trudeau<tr>jtrudeau<eoe>Paul Martin<tr>pmartin<eoe>Jean Chretien<tr><eos>},
	and the expected label is \texttt{\small<sos>jchretien<eos>}.
	
	In general, the size of a sub-problem can vary depending on the lengths of the records in source and target, the length limitation of the large language model being employed, and maybe the complexity of transformations. In our case, each example consists of two rows, a source and a target. Assuming that the input consists of $k$ examples and a source row to be transformed, and using a language model that takes 512 tokens, the length of each row is limited to $\lfloor 512/(2k+1)\rfloor$ tokens, ignoring special tokens and separators. Also, more complex transformations generally require more examples to better describe the operations. For instance, consider the example \texttt{(`john junior turner', `jturner')} in the context of the example given in Section~\ref{sec:problem_def}. With only one example, one cannot tell if the letter \texttt{j} in the target is derived from \texttt{`john'} or \texttt{`junior'}. However, with two examples, there is less chance of an ambiguity. Unless explicitly stated otherwise, we set the number of examples in our contexts to two.
	
	\subsection{Tokenizer and Model}
	\label{sec:model}
	As large language models expect vectors as inputs, the input needs to be tokenized and each token is assigned a vector. Even though tokenization is considered as the first step in many NLP pipelines, the choice of tokenization is less obvious when working with tabular data.
	Conventional NLP models (e.g., word2vec, GloVe) use the vocabulary words as the unit for tokenization, and out-of-vocabulary tokens are usually all collapsed into an unknown token. Recent deep models utilize a more efficient tokenization that may better handle the morphological structure of the words, out-of-vocabulary words, or low-resource languages. Transformer-based language models~\cite{t5,bart,gpt,gpt2} generally use either WordPiece~\cite{schuster2012japanese} or Byte Pair Encoding (BPE)~\cite{sennrich2016neural} tokenizations, where more frequent consecutive byte pairs are represented as new symbols in the vocabulary (analogous to text compression techniques). In both cases, words are broken down into subword tokens based on the frequency of each subword in the language. As a result, common words are represented as a single token whereas rare words are broken down into multiple tokens.

	However, in our problem setting, a subword-level tokenizer may not be the right choice. Subword tokenizers are mainly optimized to pick subwords of natural language or words in the input domain while in tabular data, the values can be from any domain including words not from a specific language. Understanding meaning and semantics of the words or splitting them into smaller meaningful parts is not essentially helpful in predicting the output as each character may independently contribute to the output value.
	For instance, consider the pair \texttt{(``Justin Trudeau'', ``J.Trudeau'')}, where the first character \texttt{J} in \texttt{Justin} is producing the letter \texttt{J} in \texttt{J.Trudeau}. In a different example pair, \texttt{(``Justin Trudeau'', ``Trudeau, Justin'')}, the word \texttt{Justin} is used in the output as a single token. A pretrained subword-level tokenizer may not be the best choice for such input tokenization. 
	A similar problem arises in low-resource languages that lack enough data for training a tokenizer. It is shown that character- or byte-level tokenizers work better in those settings~\cite{xue2021mt5,byt5}.
	On the same basis, we adopt byte-level tokenizer ByT5~\cite{byt5} in our work.
	Recent work has shown that byte-level models are competitive with their subword-level counterparts~\cite{byt5}, especially in tasks dealing with short-length texts. 
	Generally, table cells store short-length content and our serialization technique also generates short-length contexts with only two examples. Taking this into account, we use a byte-level UTF-8 encoder as the tokenizer, which benefits from the accuracy of character-level tokenizers and maintains a relatively short input length passed to the model.
	
	With the input represented as a sequence of tokens, the problem becomes a text-to-text transformation, where a suitable choice for the model architecture is a sequence-to-sequence model that comprises an encoder and a decoder block.
	Most table embedding models utilize WordPiece and BPE tokenizers~\cite{tacl_ttdr,dong2022table} with an encoder-only architecture or rely on table metadata and schema, rendering them unsuitable for our problem setting. An effective model should not only demonstrate adequate natural language understanding but also possess the capability to interpret and leverage patterns within tabular data. We strike this balance by opting for a general-purpose LLM for natural language knowledge and subsequently training it to excel in syntactic transformations.
	Recent models are stacking a same number of transformer~\cite{transformers} layers for both encoder and decoder. However, it is shown that when the input is a sequence of characters, using a deeper encoder containing more layers of transformers, referred to as unbalanced architecture, performs better than a balanced model~\cite{byt5}. ByT5~\cite{byt5} is a recent byte-level model with an encoder block three times deeper than the decoder block, which we use as a starting point for the training process. Unlike the original model, which masks parts of the output, we mask all characters in the target, and the training objective is to predict the masked bytes. The decoder is auto-regressive, and only the initial token, \texttt{<sos>}, is passed to the decoder.
	In the next sections, we will delve into the details of passing the input and predicting an output.

	\subsection{Aggregator}
	\label{sec:aggregator}
	We have decomposed the problem of transforming a table into a set of smaller tasks (see Section~\ref{sec:serialization}), where each of these tasks is carried out using a sequence-to-sequence model, as discussed in Section~\ref{sec:model}. To exploit all provided examples in the prediction process, each input is fed into the model multiple times, each time with a different context. If we denote the number of trials with $n$, the model will predict $n$ target candidates, denoted as $O_i = \{o_{i1}\ldots o_{in}\}$, for each row $s_i \in S$ in the source.
	
	In an ideal setting where there is no noise or inconsistency in the data and the model performs with no error, all of the predicted values for a specific source $s_i$ should be the same, i.e. $o_{i1} = o_{i2} = \ldots = o_{in}$. However, inaccurate examples, noisy input rows, and inconsistencies among multiple rows can lead to different predictions for a particular source row. It should be noted that due to the limitations in the model's input length, it is not feasible to pass the entire example set to the model, and instead, we create various subsets, each of which is treated as an independent problem. While noise in the examples may affect the output in some subsets, we ensemble the outputs generated under different contexts to obtain the best possible output. Consequently, the predicted target $t_i$ for the source $s_i$ can be estimated as
	\begin{equation}
		t_i = \underset{o_{ij} \in O_i}{\mathrm{argmax }} \text{\space} P(C_i | o_{ij}),
	\end{equation}
	where $C_i \subseteq C$ is a subset of contexts that may include example sets that are relevant to source $s_i$, and $C_i$ may also be limited in size, for example to $n$. By applying Bayes' theorem, we have
	\[
	P(C_i | o_{ij}) = \frac{P(o_{ij} | C_i) P(C_i) }{P(o_{ij})}.
	\]
	Assuming a uniform prior probability $P(o_{ij})$ for the predictions and treating $P(C_i)$ the same for all predictions, these terms can be ignored, and $P(C_i | o_{ij}) \propto P(o_{ij} | C_i)$ can be used as a proxy for finding the argmax. Also, assuming independence among predictions, it is possible to use the maximum likelihood estimation to calculate $P(o_{ij} | C_i)$, i.e.
	\begin{equation}
		t_i = \underset{o_{ij} \in O_i}{\mathrm{argmax }} \text{\space} P(o_{ij} | C_i) \propto \frac{|o_{ij}|}{|O_i|}
	\end{equation}
	where $|o_{ij}|$ is the frequency of $o_{ij}$ in $O_i$ and $|O_i|$ is number of possible predictions.
	
	\subsection{Downstream Tasks}
	\label{sec:tasks}
	Given a source row and a set of source-target example pairs, the proposed model generates a target row following the examples. This framework can be useful in many downstream tasks such as auto-completion and auto-filling spreadsheets~\cite{FlashFill,BlinkFill}, predicting missing values, error correction~\cite{he2016interactive}, and joining~\cite{icde,autojoin}. In this section, we review the particular task of joining heterogeneous tables and leave generalizing the framework for other downstream tasks for the future works.

	Consider a join scenario where a source table $S$ and a target table $T$ must be joined, based on some columns that are not formatted the same but there is a mapping from source to target. Examples of mappings may be provided by the user or obtained automatically~\cite{icde}. The model can be invoked to generate a target $f(s_i)$ for each source row $s_i$, utilizing the examples as discussed earlier.
	Unlike the case for some other tasks such as filling missing values where an exact prediction of the missing value is needed, an exact prediction is not necessary for a join.
	Instead, the goal is to use $f(s_i)$ as a bridge to establish a connection between rows in $S$ and $T$. For instance, under the setting where each value in the source is matched with a single value in the target (e.g., a primary-foreign key relationship), one needs to find the closest match in $T$ for $f(s_i)$. 
	This allows small discrepancies between predicted and target values, without affecting the join. 
	There is a great chance that a significant string similarity exists between a model predicted value $f(s_i)$ and the corresponding value in $T$. In many cases, this similarity is enough to perform the join. Therefore, for each $(s_i, f(s_i))$ pair, we can select $t_j \in T$ such that it yields the minimum edit distance between the two strings. This can be formalized as follows:
	\begin{equation}
		m_i = \underset{t_j \in T}{\mathrm{argmin }} \text{\space} edit\_dist(f(s_i), t_j) 
	\end{equation}
	where $m_i$ is considered a match in the target for $s_i$. The approach can be generalized to cases where a value in the source is matched with either no values or multiple values in the target. To allow such many-to-many joins, one may set lower and upper bounds for the edit distance instead of aiming for the minimum distance.
	
	\section{Experiments and analysis}
	In this section, we evaluate our proposed model and analyze its performance under different settings. We also discuss our training data generation and the process of training our model.
	
	\subsection{Dataset for Training DTT}
	\label{sec:inp_data}
	Pretrained language models are generally trained on large text corpora, and a common approach for training them involves  masking out a word and predicting it, as seen in popular models such as T5~\cite{t5}, BERT~\cite{bert} and GPT-2~\cite{gpt2}. By using this approach, large amounts of unlabeled data can be utilized to train the model. Nonetheless, our particular task requires a vast set of source and target examples, grouped according to the transformations that map source examples to their corresponding targets. To the best of our knowledge, such a dataset is not currently available, and our experiments have shown that even advanced generative models pretrained on natural language text, such as T5 and GPT-2, are not capable of performing this task without extensive fine-tuning and training. This is because entries in real-world tables are typically short and have little relevance to other entries in the same column, aside from sharing the same domain (e.g., individual names). As a result, the prior language knowledge of these general models is less likely to be useful for this task.
	To address this challenge, we propose generating synthetic data to train the model. Before delving into the details of data generation, however, it is important to first review the desired features of the training data.
	
	\subsubsection{Training data features}
	The training data must possess several key features.
	First, it should be organized as source-target pairs, categorized by their corresponding transformations, as previously discussed. It is worth noting that the mapping function can be general, as the model does not need to know the function itself; rather, it only requires the output of the function for the source examples and that the generated examples by the same mapping are grouped together.
	Second, the dataset must be sufficiently large to train a model with hundreds of millions of parameters, which is typical for many language models.
	Third, the dataset should cover a broad range of textual transformation patterns and various input lengths.
	Finally, the generated data should not be limited to the words in any specific language since many terms 
	in table cells are not dictionary words and may not be limited to a particular language.

	Overall, the primary purpose of training data in our case is guiding the model to understand the mapping corresponding to a set of source-target example pairs. In this context, different combinations of edit operations can be performed on the source examples to generate the target outputs. Unlike NLP models that rely on understanding the syntax and the semantics of input, our model primarily focuses on discovering textual patterns and string operations. Hence,  character-level tokenization is preferred in our case. In the rest of this section, we will delve into the process of generating a synthetic dataset to train our model.

	\subsubsection{Training data generation}
	To generate our synthetic dataset we first build a set of textual transformations, denoted as $T$, each consisting of a sequence of basic transformation units. 
	We use the basic transformation units in recent literature~\cite{autojoin,icde}, which include \texttt{substring}, \texttt{split}, \texttt{lowercase}, \texttt{uppercase}, and \texttt{literal}.  These units have their natural meanings: \texttt{substr} selects a portion of the input based on start and end parameters, \texttt{split} breaks the input by a given  character and selects one part, \texttt{literal} returns a constant, and \texttt{lowercase} and \texttt{uppercase} return the lowercase and uppercase forms of the input respectively. Each unit copies either parts of the input or a literal to the output, and the output of a transformation is the concatenation of the outputs of its units.
	We randomly choose the units, parameters, and the length of each transformation in terms of the number of units, to provide a diverse set of examples. While the aforementioned transformations are expected to cover many cases of transformations in real-world settings~\cite{autojoin, icde}, our aim is not to limit the model to learning a fixed set of transformations. Our findings indicate that, with sufficient independent training examples, the model can learn to find any necessary transformation even with a limited set of pre-defined transformations.  The construction of transformations mainly helps us group input examples that follow a same mapping, but the model is not aware of the transformations and uses regular cross-entropy loss at the character level to learn a mapping that transforms the source into the target. 
	
	The transformations in Nobari~et~al.~\cite{icde} and Zhu~et~al.~\cite{autojoin} do not allow stacking of the units where one unit is applied on top of another unit. For the same reason, they introduce complex transformation units such as \texttt{splitsubsting} which stacks substring on top of split, with the output of one operation fed to the other. 
	Instead of introducing many such new units, we allow random stacking of up to three transformation units. The stacking here refers to passing the output of one transformation unit to another one. Since our units include lower case and upper case transformations, the case of input may change in some transformations and not others. 
	
	For each transformation $tr \in T$, a set of examples is generated. To create these examples, a source text is randomly generated consisting of a mix of alphabetic and numeric characters, symbols, and special characters. The length of the input is selected at random. The transformation $tr$ is then applied to source texts to generate a set of examples, denoted as $I_{tr} = \{ (s_i, t_i)\}_{1 \leq i \leq u}$. Using random text instead of dictionary words avoids any potential bias towards natural language words and grammatical structures. To form example sets, subsets of size 3 are selected from $I_{tr}$. Each example set is then serialized, as discussed in Section~\ref{sec:serialization}, with the target of the last example masked and labeled as the target for use in forming context sets for model training.

	\subsection{Dataset for Evaluation}
	To evaluate the effectiveness of our approach and compare its performance with state-of-the-art baselines, we use three real-world datasets as well as four synthetic datasets. In what follows, we provide a detailed explanation of each dataset.
	
	\noindent
	\textbf{Web Tables Dataset (WT)} This benchmark was initially introduced by Zhu~et~al.~\cite{autojoin} and was also used as a benchmark in Nobari~et~al.~\cite{icde}. The dataset includes 31 pairs of tables from 17 distinct topics, with an average of 92.13 rows per table and an average length of 31 characters per input source. The tables were sampled from Google Fusion tables by identifying tables that appeared in the results of the same queries but were formatted differently. The rows were then manually annotated by the authors. This benchmark contains natural noise and inconsistencies, and not all entities can be transformed using traditional string-based transformations, which makes this dataset a relatively challenging benchmark~\cite{icde}.
	
	\noindent
	\textbf{Spreadsheet Dataset (SS)} This dataset includes 108 pairs of tables, sourced from Microsoft Excel product team and user help forums, specifically focused on users' data cleaning issues. The tables are comprised of spreadsheet pages that present the same information in different formats. The dataset encompasses the public benchmarks presented in FlashFill~\cite{FlashFill} and BlinkFill~\cite{BlinkFill}, and was published in 2016 Syntax-Guided Synthesis Competition (SyGuS-Comp)~\cite{sygus2016}. On average, each table in the dataset contains 34.43 rows and 19 characters per input source. Compared to web tables, this dataset features considerably less noise and inconsistency. 
	
	\noindent
	\textbf{Knowledge Base Web Tables (KBWT)} This dataset, introduced by Abedjan~et~al.~\cite{DataXFormer}, primarily comprises tables sourced from a Knowledge Base (KB), requiring semantic transformations and additional information from the KB. The ground truth is manually annotated by the authors. This is notably different than the WT benchmark, which is predominantly centered around textual transformations. We chose single-column tasks from this dataset for our evaluation, comprising 81 pairs of tables. On average, each table contains 113 rows and 13 characters per input source. 
	
	\noindent
	\textbf{General Synthetic Dataset (Syn)} This is a synthetic dataset that contains 10 table pairs. Each pair is generated by applying a randomly generated textual transformation to a set of random input sources to create the output table. The transformations are constructed by putting together a random sequence of 3 to 6 units, the same as those discussed in Section~\ref{sec:inp_data}, with random parameter sets. Unless stated differently, the dataset contains 10 tables, each of which contains 100 rows. Input length is randomly chosen in the range of 8 to 35, and no artificial noise is added to the dataset. While the model has been exposed to the units during the training, the transformations, the parameter sets of the units, and the inputs are unseen during the training process. 
	
	The next three synthetic datasets introduce various levels of difficulty in transformations. As our model, akin to many LLMs, operates as an auto-regressive model predicting each output token based on context and previous tokens, it is anticipated that as the need for a greater number of edit operations increases for any given input, the prediction task becomes  more challenging for the model. We utilize this notion as a heuristic to determine the difficulty level of our synthetic datasets.

	\noindent
	\textbf{Easy Synthetic Dataset (Syn-RP)} This is a synthetic dataset containing 5 pairs of tables. Each pair is formed by randomly replacing one character with another (for example, the character `/' might be replaced with `-' for all rows). This dataset resembles simple formatting changes such as replacing a slash in a phone number with a hyphen. This replacement operation is not a transformation unit that exists in the model's training data and is thus unseen by the trained model.
	Each table contains 50 rows, and the length of input sources is randomly selected from a range of 8 to 35, unless stated otherwise. Since our model is generating the output character-by-character, we measure the difficulty of datasets based on the number of required edit operations. Accordingly, this is an easy dataset considering that only a few characters in the input need to be changed to generate the desired output.
	
	\noindent
	\textbf{Medium Synthetic Dataset (Syn-ST)} This synthetic dataset is similar to the previous one in terms of the number of table pairs and the input length. Each table pair is constructed by applying a single substring transformation unit to the input, with the start and end parameters selected randomly. Substring is one of the units included in the model's training data. In terms of difficulty, this dataset is considered to be medium-level based on the number of edit operations required.
	
	\noindent
	\textbf{Difficult Synthetic Dataset (Syn-RV)} This synthetic dataset consists of 5 tables, each containing 50 rows with input sources randomly selected to have a length between 8 to 35 characters. In this dataset, the target output is obtained by reversing all characters in the source (for instance, ``Hello'' is changed to ``olleH''). This benchmark is considered difficult since almost all characters in the input source must be changed to generate the expected target, and the model has not seen any such transformation in training.

	\subsection{Experimental Setup}
	Our model, DTT, was trained on a synthetic dataset containing 2000 groupings of transformations, each corresponding to a transformation, as discussed in Section~\ref{sec:inp_data}. For each grouping, we generated 10 pairs of source-target samples with randomly chosen input lengths ranging from 8 to 35. 
	80\% of the samples were used for training and the other 20\% were the validation set. We also conducted experiments with other sample sizes and input lengths for training the model, and the results are discussed in Section~\ref{sec:varySampleSize}.
	
	To evaluate the performance of our model, we divided the rows of each input table in our datasets into two equal-sized sets, denoted as $S_e$ and $S_t$. The former provided context examples to be passed to the model, while the latter was used for testing. Since DTT is an example-driven method, the selection of these examples is critical to the model's performance. To ensure the robustness of our predictions, we employ a technique where each input is fed to the model five times, and each time a distinct set of randomly chosen examples from $S_e$ is given as context. The results of those trials are aggregated, as discussed in Section~\ref{sec:aggregator}, to produce a final prediction.

	\subsection{Evaluation Metrics}
	We evaluate the performance of our models based on precision, recall, and F1-Score. This evaluation is in the context of heterogeneous join, as discussed in Section~\ref{sec:tasks}, where for a given source-target sample $(s,t)$, we consider a model prediction correct if it has the minimum edit distance with the target $t$. In our case, precision represents the fraction of correct predictions that join with the target, recall measures the fraction of source rows that are correctly mapped, and F1-score is the harmonic mean of the two. It is important to note that not all source rows may be mapped due to various reasons\footnote{For AFJ, a threshold for similarity distance is set and based on that threshold, some source rows will not have a match. In CST, a match may not still be found after applying the detected transformations to all input rows. In our approach, the language models may just return <eos> with no prediction.}. 
	In addition to the above metrics, we also report the Average Edit Distance(AED) and Average Normalized Edit Distance (ANED), which indicates the extent to which a prediction may differ from the given target. The normalization is performed based on the target length, enabling comparability across different datasets and lengths.
	All reported metrics for each dataset are the average over all tables in the dataset.

	\subsection{Performance Compared to Heterogeneous Join Baselines}
	In this section, we evaluate the performance of our model on the end-to-end task of heterogeneous or unequal table join. The task simulates the scenario where source and target columns are in two different tables that need to be joined.
	To provide a point of reference, we compare the performance of our model to three current state-of-the-art baselines:
	Common String-based Transformer (CST)~\cite{icde}, Auto-FuzzyJoin (AFJ)~\cite{auto-fuzzy-join}, and Ditto~\cite{DITTO}. CST finds a set of textual transformations given a set of examples to transform tables for joinability and AFJ uses a set of similarity functions to detect the most probable rows to be joined. Ditto is an entity matcher, which fine-tunes pre-trained large language models (DistilBERT~\cite{sanh2019distilbert} in our experiments) for this task.

	Table~\ref{tab:joinres} summarizes the performance of DTT and the baselines, in terms of precision, recall, and F1-score, denoted as P, R, and F respectively.  
	The results show that DTT outperforms the baselines on all real-world datasets in terms of F1-Score and recall. On the synthetic datasets, our approach outperforms the baselines on three out of four datasets. On \textit{Syn-RP} and \textit{Syn-ST} datasets, our approach is either comparable or slightly worse than the baselines. 
	The reason is that these datasets are relatively easy, with a significant textual similarity between the source and target. CST exhaustively searches the space for substring transformation, which is the only transformation used in the \textit{Syn-ST} dataset. Moreover, AFJ is based on the textual similarity, and every target in \textit{Syn-ST} is a substring of the source, leading to a significant similarity between source and target. Similarly, Ditto is fine-tuned to observe similarities between source and target and match them based on that similarity. Therefore, these datasets favor the baselines. Nevertheless, DTT still achieves an F1-score of 88\% on the \textit{Syn-ST} dataset and a perfect F-score of 100\% on \textit{Syn-RP}, which is equal to AFJ in performance and better than CST and Ditto. Moreover, our observations indicate that when there exists a significant textual or semantic similarity between rows in the target, Ditto and AFJ (which are based on matching) may end up with many false positives, while DTT performs better in such cases as the target text is generated independent of the similarity among rows in target.

	\begin{table*}[tbp]
		\centering
		\setlength{\tabcolsep}{3.8pt}
		\caption{Performance compared to heterogeneous join baselines}
		\Small
		\begin{tabular}            {|l||c|c|c|c|c||c|c|c||c|c|c||c|c|c|}
			\toprule
			& \multicolumn{5}{c||}{Our Approach (DTT)}    & \multicolumn{3}{c||}{CST} & \multicolumn{3}{c||}{AFJ} & \multicolumn{3}{c|}{Ditto} \\
			\midrule
			Dataset & P     & R     & F     & AED   & ANED  & P     & R     & F     & P     & R     & F     & P     & R     & F \\
			\midrule
			WT    & 0.951 & 0.950 & \textbf{0.950} & 6.155 & 0.232 & 0.879 & 0.726 & 0.713 & 0.935 & 0.672 & 0.708 & 0.852 & 0.785 & 0.721 \\
			SS    & 0.954 & 0.952 & \textbf{0.953} & 2.399 & 0.135 & 0.995 & 0.792 & 0.812 & 0.943 & 0.662 & 0.691 & 0.745 & 0.789 & 0.663 \\
			KBWT   & 0.276 & 0.248 & \textbf{0.254} & 11.180 & 0.786 & 0.962 & 0.081 & 0.083 & 0.749 & 0.067 & 0.093 & 0.276 & 0.228 & 0.131 \\
			Syn   & 0.934 & 0.934 & \textbf{0.934} & 6.986 & 0.150 & 0.990 & 0.259 & 0.324 & 0.993 & 0.490 & 0.511 & 0.317 & 0.854 & 0.274 \\
			Syn-RP & 1.000 & 1.000 & \textbf{1.000} & 0.816 & 0.027 & 1.000 & 0.816 & 0.897 & 1.000 & 1.000 & \textbf{1.000} & 0.801 & 1.000 & 0.875 \\
			Syn-ST & 0.880 & 0.880 & 0.880 & 5.032 & 0.316 & 1.000 & 1.000 & \textbf{1.000} & 1.000 & 1.000 & \textbf{1.000} & 0.830 & 1.000 & 0.898 \\
			Syn-RV & 0.632 & 0.632 & \textbf{0.632} & 33.600 & 0.852 & 1.000 & 0.000 & 0.000 & 0.990 & 0.020 & 0.037 & 0.495 & 0.160 & 0.234 \\
			\bottomrule
		\end{tabular}%
		\label{tab:joinres}%
	\end{table*}%
	
	There are significant differences between DTT and the baselines.
	CST is limited in its ability to extract transformations, and cannot perform a join when there is no clear copying relationship between the source and target, as is the case with the \textit{Syn-RV} dataset where the target is obtained by reversing the input. As a result, CST achieves a 0\% F1-score on this dataset.
	AFJ and Ditto, on the other hand, employ similarity functions to determine if source and target values can be joined. However, this approach struggles when there is not much similarity between the source and target, as demonstrated by its performance on the \textit{Syn-RV} dataset. Such challenges are common in real-world data.
	DTT, in contrast, leverages the provided examples to generate the desired output without relying on textual similarity or being bounded by the length of transformations. Hence, DTT demonstrates notably superior performance compared to the baselines, particularly 
	on more challenging datasets, including the real-world \textit{WT} and \textit{KBWT} datasets, as well as the synthetic \textit{Syn} and \textit{Syn-RV} datasets.
	For instance, DTT outperforms the baselines by a large margin on \textit{Syn-RV}, where the target is obtained by reversing the order of characters in the input.

	Two more interesting observations can be made here. Firstly, to achieve a good performance on the join, it is not necessary to predict every single character correctly. Our framework can tolerate inaccuracies by aggregating results from multiple examples and using an edit-distance-based metric to form join candidates. For example, in \textit{Syn-RV} dataset, while the average normalized edit distance is more than 80\%, the F1-score for join prediction is 63\%. Secondly, our model performs very well on all real-world datasets and two synthetic datasets \textit{Syn-RP} and \textit{Syn-RV}, despite the fact that our training data did not include any operation that simulates reversing the input or replacing a character, and no real-world examples or transformations were included in the training data. 
	This highlights that the model is not limited to a given set of transformations, but rather focuses on extracting character-level patterns from the given set of input examples. Furthermore, while the model is only fine-tuned on textual transformations, our experiments suggest that it can cover some semantic transformations that require information from a knowledge base because of its prior knowledge of natural language and web data. More specifically, on the \textit{KBWT} dataset, where tables require transformations that leverage information from a knowledge base, DTT achieves an F1 score of 0.25 across all tables, surpassing all baseline models. 
	
	As an additional baseline for the KBWT benchmark, we also consider DataXFormer~\cite{DataXFormer}. We compare DTT with their unsupervised model, given that our model has not encountered any transformations from the benchmarks during the training process and is exclusively trained on synthetic textual transformations. Despite DataXFormer being optimized for relationships and tables in KBs, our approach performs on par with it. 
Our observations indicate that DTT excels in scenarios where general knowledge is crucial for transformations, as evident in tasks like \textit{`Country To Citizen'} and \textit{`State To Abbreviation'}. However, in tasks such as \textit{`ISBN To Author'} and \textit{`City To Zip'}, where more parametric knowledge from a KB is essential, the model does not perform at the same level as observed  in the former tasks.
This limitation can be mitigated by leveraging general-purpose LLMs, as elaborated in Section~\ref{sec-multi-model}.

Finally, in terms of comparing the runtime of DTT and our baselines, a direct comparison is not possible since DTT and Ditto require a GPU architecture whereas CST and AFJ require CPU and memory. That said, some observations can be made on the scalability of the models.
The time required to predict a mapping for each row in DTT is independent of the number of rows and grows linearly with the length of the rows, whereas this time grows quadratically with the number of rows and polynomially with the length in CST (see \cite{icde} for details).
While the edit distance calculation in the joining process depends on the number of rows, our experiments suggest that the growth in the runtime of DTT is noticeably less than that of CST when input length increases. 
For instance, with our machine setup\footnote{Our experiments were conducted on a machine with Nvidia RTX 3090 GPU and AMD EPYC 7601 CPU with 64GB RAM.}, processing a table with row length set to 5 characters from our synthetic dataset takes 5 seconds for DTT and 3 seconds for CST.
However, when the input length increases to 50 characters, DTT needs less than 17 seconds, while CST takes around 90 seconds to complete the join. It should be noted that the runtimes reported for DTT are the summation of decomposition, all 5 trials, and the aggregation time. 
For scalability in terms of the number of rows, we compared their performance on two tables from our spreadsheet dataset, ``phone-10-short'' and ``phone-10-long'', both with an average of 17 characters per row. The former has 7 rows, while the latter has 100. DTT takes 3 and 22 seconds respectively for short and long tables, while the same experiments require 4 and 366 seconds for CST, 4 and 38 seconds for AFJ, and 1, 10 seconds for Ditto. It should be noted that while Ditto needs less running time (due to its smaller model), when the input length increases, the growth is more comparable to DTT. Also, DTT outperforms Ditto in terms of accuracy. This indicates how our framework scales better in terms of runtime when the input grows either horizontally or vertically.

\subsection{Performance Compared to Large Language Model Baselines} 
Large Language Models (LLM) can be employed in many downstream tasks including joining heterogeneous tables. It has been shown that the recent models perform relatively well under zero- or few-shot settings~\cite{chen2022large,gpt3}, hence they set a strong baseline. In this section, we compare the performance of our model to GPT-3~\cite{gpt3}, a state-of-the-art LLM (at the time of conducting our experiments) with exceptional performance on many tasks. Compared to our ByT5-base~\cite{byt5} model that is fine-tuned on only 20,000 synthetically-generated samples and contains near 582M parameters, GPT-3 models are trained on billions of documents and resources such as web tables and have at least one to two orders of magnitude more parameters.
Our experiment with GPT-3 is under few-shot setting with 1, 2, 3, 5, and 10 randomly selected samples from each table given as examples.
Zero-shot setting is not applicable in our case since an input source row can be mapped to an unlimited number of targets without any examples. 

\begin{figure*}[tbp]
	\centering
	\includegraphics[width=1\linewidth]{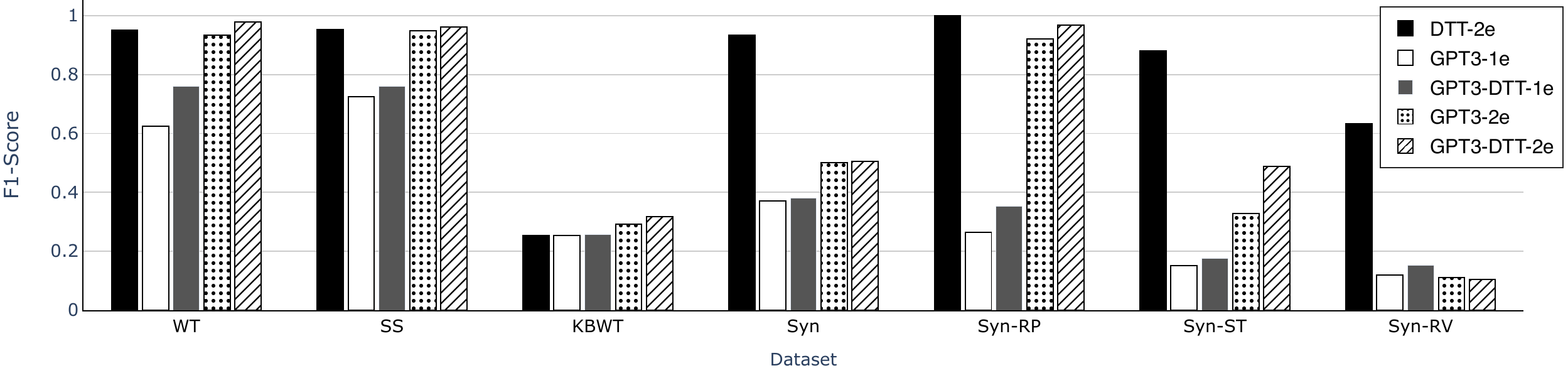}
	\caption{Performance compared to GPT-3 as well as the performance of the combined model}
	\label{fig:gpt3}
\end{figure*}

At the time of this writing, GPT-3 models are not published publicly and are only accessible through OpenAI commercial API\footnote{https://openai.com}. We use the Curie model of GPT-3. Curie is claimed to be extremely powerful, fast, and capable of many advanced tasks\footnote{Based on platform documentations on openai.com}. Nevertheless, the model specification and the number of parameters are not publicly announced. 
Comparing the general performance of the Curie model with the performance reported for various sizes of GPT-3~\cite{gpt3}, it can be assumed that the Curie model has about 7B parameters
and is trained on a huge text data from different datasets.

We ran two sets of experiments to analyze the performance of GPT-3 for unequal join. First, as the common method of using LLMs, we passed our examples as an input sequence to GPT-3 and considered the model output as the expected target. The serialization used for GPT-3 was the same as DTT, as discussed in Section~\ref{sec:serialization}. 
%However, we do not limit the number of examples to two and all of them are concatenated. 
In the second experiment, we used GPT-3 as a replacement for our fine-tuned ByT5 model (and byte-level tokenizer) inside our framework, keeping the serializer and aggregator from DTT.

As shown in Figure~\ref{fig:gpt3}, GPT-3 struggles on the task with just one example despite some recent work suggesting that LLMs are capable of one-shot table reasoning~\cite{chen2022large}. However, providing two examples significantly boosts its performance, especially on real-world data, bringing it on par with DTT. 
In our synthetic datasets, however, DTT performs significantly better than GPT-3. The lack of publicly available data on GPT-3 Curie model's size and specification makes it challenging to conduct a more in-depth comparison.
It can be noted that GPT-3 is trained on numerous web resources, including web tables, which increases the likelihood that it has encountered various representations of common entities and tables on the web. Since our real-world datasets are gathered from tables on web pages, this could explain why the model performs better on real-world datasets than on synthetic datasets, especially on \textit{KBWT} benchmark, where external information from a KB is required for an accurate transformation. Conversely, synthetic datasets consist of sequences of random characters that may not be tokens from natural language, and GPT-3 may not have encountered them during its training. Consequently, its performance on most of the synthetic datasets is weak and, in some cases, significantly inferior to DTT, especially in the \textit{Syn-RV} dataset, where the target and source are substantially different.
Our ByT5-based model, however, is trained to extract patterns among a sequence of characters, allowing it to perform better on more challenging synthetic datasets.

In our second set of experiments with GPT-3, we used our framework and replaced the LLM module with GPT-3. By default, our framework employs two context example pairs because ByT5 has a maximum limit of 512 character-level tokens, and if a longer sequence is given to the model, it will be truncated. However, the limit in GPT-3 Curie model is 2048 subword-level tokens. This allows us to increase the number of example pairs that are given to the model. In our experiment with GPT-3 integrated into the DTT framework, we varied the number of examples from one to five. 
As demonstrated in Figure~\ref{fig:gpt3} and Table~\ref{tab:gpt3}, using GPT-3 within our framework boosts its performance, in terms of both the F1-score and ANED, on nearly all datasets when the same number of examples were provided. For instance, the average F1-score of GPT-3 across all datasets increases from 0.577 to 0.618 with two examples and from 0.675 to 0.703 with five examples when integrated into the DTT framework. This demonstrates how our decomposition and aggregation steps can be used with other larger models to gain a performance boost.

\begin{table*}[tbp]
	\centering
	\setlength{\tabcolsep}{2.4pt}
	\caption{Performance of GPT-3 as well as that of the combined model}
	\scriptsize
	\begin{tabular}{|l||c|c||c|c||c|c||c|c||c|c||c|c||c|c||c|c|}
		\toprule
		& \multicolumn{2}{c||}{GPT3-1e} & \multicolumn{2}{c||}{GPT3-2e} & \multicolumn{2}{c||}{GPT3-3e} & \multicolumn{2}{c||}{GPT3-5e} & \multicolumn{2}{c||}{GPT3-DTT-1e} & \multicolumn{2}{c||}{GPT3-DTT-2e} & \multicolumn{2}{c||}{GPT3-DTT-3e} & \multicolumn{2}{c|}{GPT3-DTT-5e} \\
		\midrule
		Dataset & F     & ANED  & F     & ANED  & F     & ANED  & F     & ANED  & F     & ANED  & F     & ANED  & F     & ANED  & F     & ANED \\
		\midrule
		WT    & 0.625 & 0.499 & 0.933 & 0.151 & 0.954 & 0.108 & 0.966 & 0.088 & 0.759 & 0.341 & 0.979 & 0.072 & 0.985 & 0.074 & 0.987 & 0.073 \\
		SS    & 0.724 & 0.533 & 0.949 & 0.128 & 0.973 & 0.094 & 0.968 & 0.079 & 0.760 & 0.483 & 0.960 & 0.113 & 0.973 & 0.079 & 0.982 & 0.056 \\
		KBWT   & 0.253 & 0.858 & 0.293 & 0.769 & 0.335 & 0.710 & 0.318 & 0.701 & 0.258 & 0.848 & 0.318 & 0.729 & 0.348 & 0.700 & 0.357 & 0.664 \\
		Syn   & 0.372 & 0.889 & 0.502 & 0.619 & 0.528 & 0.522 & 0.614 & 0.418 & 0.380 & 0.902 & 0.506 & 0.567 & 0.552 & 0.495 & 0.720 & 0.387 \\
		Syn-RP & 0.264 & 0.824 & 0.920 & 0.195 & 0.976 & 0.127 & 0.984 & 0.111 & 0.352 & 0.748 & 0.968 & 0.125 & 1.000 & 0.098 & 1.000 & 0.095 \\
		Syn-ST & 0.152 & 0.941 & 0.328 & 0.812 & 0.464 & 0.726 & 0.728 & 0.527 & 0.176 & 0.923 & 0.488 & 0.717 & 0.736 & 0.589 & 0.728 & 0.510 \\
		Syn-RV & 0.120 & 0.947 & 0.112 & 0.944 & 0.112 & 0.944 & 0.144 & 0.940 & 0.152 & 0.944 & 0.104 & 0.948 & 0.120 & 0.944 & 0.146 & 0.939 \\
		\bottomrule
	\end{tabular}%
	\label{tab:gpt3}%
\end{table*}%

\subsection{Performance of multi-model DTT+GPT-3 compared to individual models}
\label{sec-multi-model}

DTT, equipped with decomposer and aggregator modules, is meticulously designed to maintain the consistency of the generated output  (\S~\ref{sec:serialization}).  Nonetheless, the flexibility of the framework allows for the utilization of multiple models, particularly when a single model proves insufficient to encompass the variations present in input tables. To explore this aspect, we conducted an experiment employing both our fined-tuned ByT5-based model and GTP-3 as a general purpose LLM. Each model underwent five equally-weighted trials, to ensure self-consistency, resulting in a total of 10 trials, all of which were subsequently passed to the aggregator module.

Table~\ref{tab:multimodel} provides a summary of the framework's performance under three settings: (1) employing only our ByT5-based model (DTT column), (2) utilizing only the GPT-3 model within the DTT framework to further improve its performance (GPT3 column), and (3) combining our model and GPT-3 as discussed (DTT + GPT3 column).
As expected, our model excels in textual transformations, while GPT-3 demonstrates better performance in transformations involving semantics or a KB. With two models, the aggregator module selects the output of the more accurate model, owing to higher consistency in the predictions of that model. Consequently, the overall performance in the combined setting closely aligns with the superior model in almost all benchmarks, and on average, the performance on the combined settings surpasses that of both single-model settings.
Interestingly, for some tables (mostly within the \textit{SS} dataset), the combined aggregator outperforms both individual models. This improvement is primarily observed when each model predicts the correct output in some trials but lacks the consistency needed to be reliably selected by the aggregator. However, when both models generate the same output, the aggregator is more likely to select that output, achieving higher consistency and ultimately a better accuracy.

\subsection{Performance Varying the Number and Length of Training Samples}
\label{sec:varySampleSize}

\begin{table}[tbp]
	\centering
	\caption{Performance of the framework with multi-model aggregator}
	\begin{tabular}{l||c|c||c|c||c|c}
		& \multicolumn{2}{c||}{DTT} & \multicolumn{2}{c||}{GPT3} & \multicolumn{2}{c}{DTT + GPT3} \\
		\midrule
		Dataset & F     & ANED  & F     & ANED  & F     & ANED \\
		\midrule
		WT    & 0.950 & 0.232 & 0.979 & 0.072 & 0.969 & 0.124 \\
		SS    & 0.953 & 0.135 & 0.960 & 0.113 & 0.982 & 0.039 \\
		KBWT   & 0.254 & 0.786 & 0.318 & 0.729 & 0.300 & 0.769 \\
		Syn   & 0.934 & 0.150 & 0.506 & 0.567 & 0.906 & 0.160 \\
		Syn-RP & 1.000 & 0.027 & 0.968 & 0.125 & 1.000 & 0.042 \\
		Syn-ST & 0.880 & 0.316 & 0.488 & 0.717 & 0.866 & 0.359 \\
		Syn-RV & 0.632 & 0.852 & 0.104 & 0.948 & 0.680 & 0.845 \\
		\midrule
		Average & 0.800 & 0.357 & 0.618 & 0.467 & 0.815 & 0.334 \\
	\end{tabular}%
	\label{tab:multimodel}%
\end{table}%

Our trained model has two important parameters: the number of samples and their length. To gain a deeper insight into the relationship between these parameters and the model's performance, we conducted an experiment where we varied the number of training samples from 0 to 10,000. Each sample here is a grouping of transformations that consists of 10 source-target pairs, and we kept the sequence length consistent with our other experiments, ranging between 8 and 35.
When the number of samples was set to zero, the ByT5 model did not undergo any fine-tuning.

\begin{figure*}
	\begin{subfigure}[b]{0.495\textwidth}
		\includegraphics[width=\textwidth]{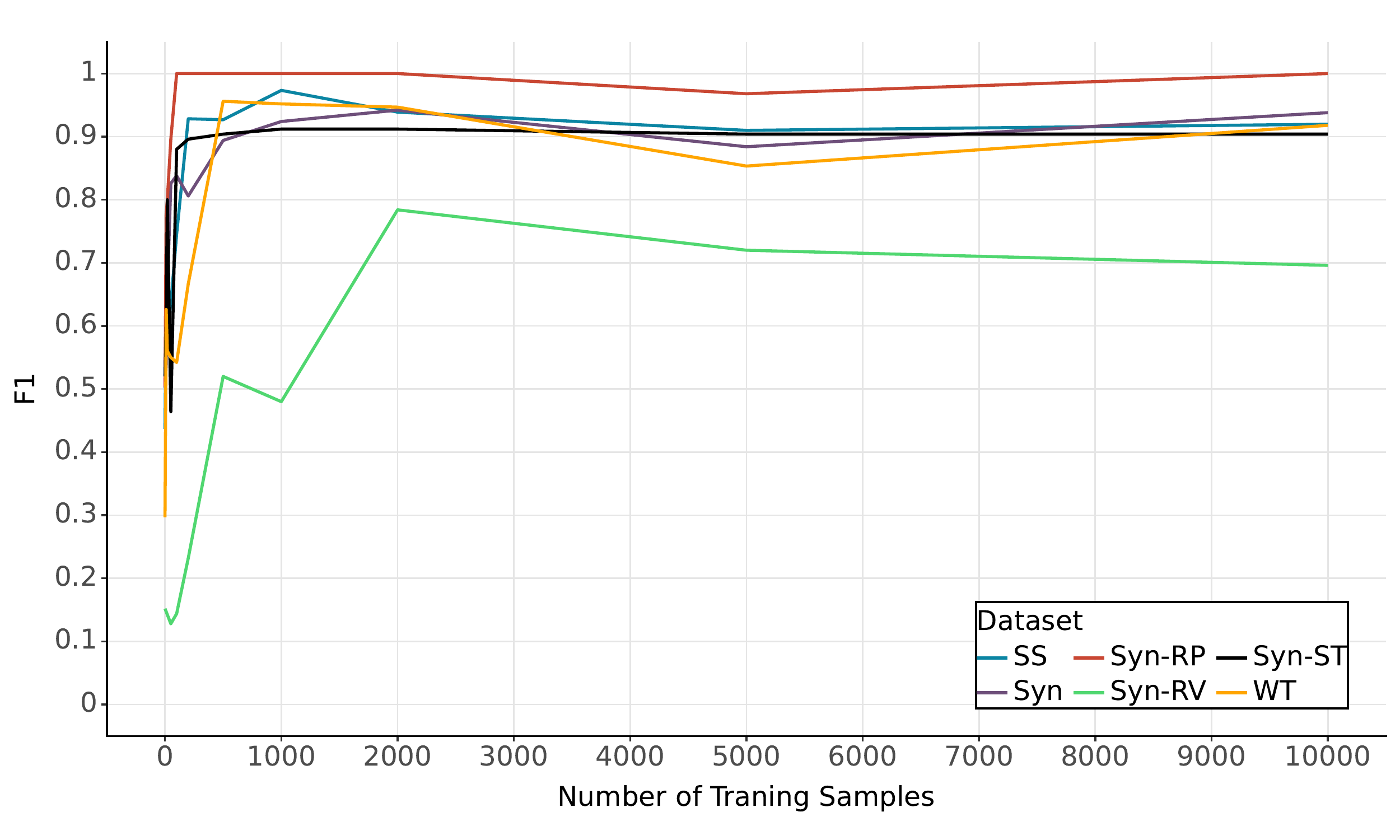}
		\caption[]{F1 score for models trained on shorter-length data } % <---
		\label{subfig:trsamp_F1_8_35}
	\end{subfigure}
	\hfill
	\begin{subfigure}[b]{0.495\textwidth}
		\includegraphics[width=\textwidth]{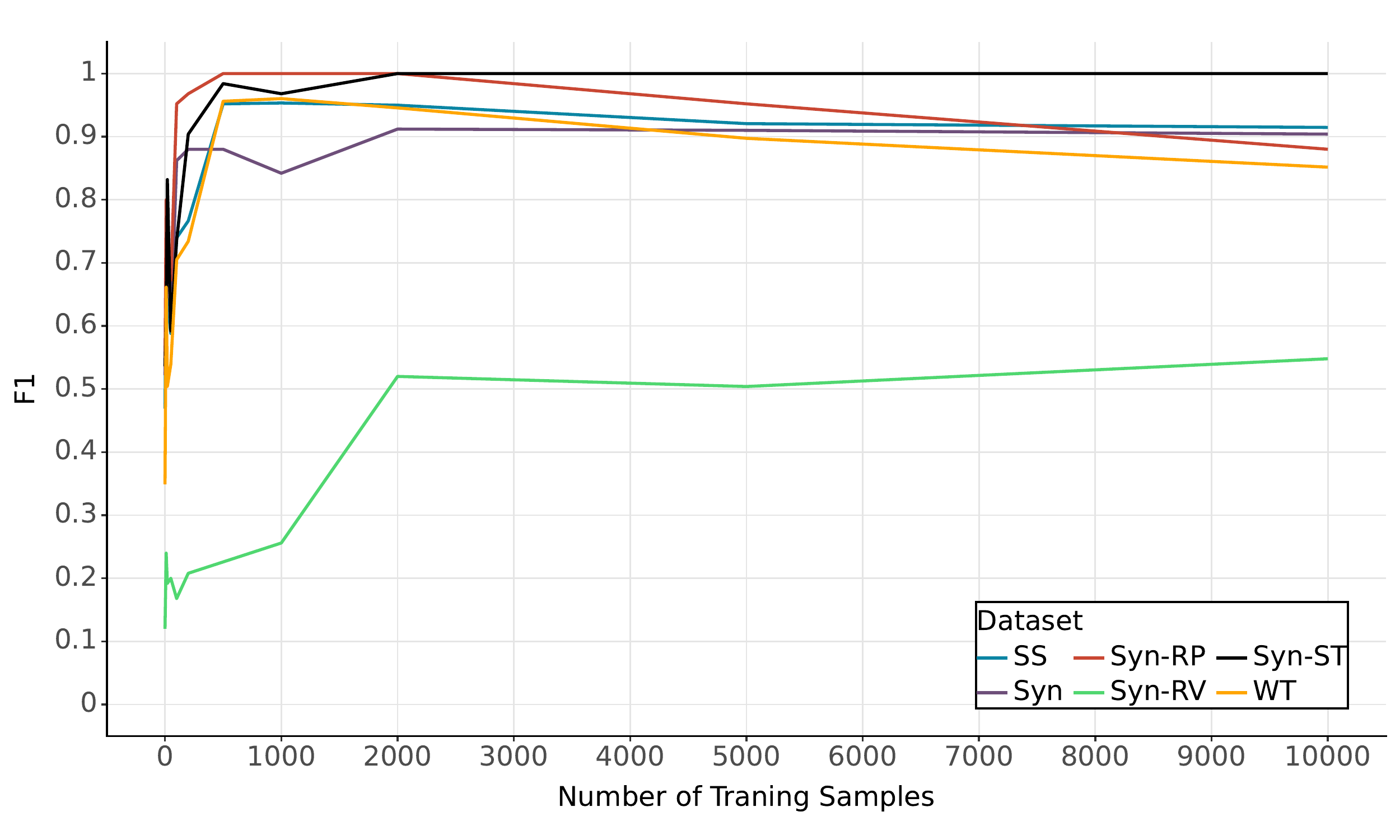}
		\caption[]{F1 score for models trained on longer-length data} % <---
		\label{subfig:trsamp_F1_5_60}
	\end{subfigure}
	
	\vskip\baselineskip
	% \vskip\baselineskip
	\begin{subfigure}[b]{0.495\textwidth}
		\includegraphics[width=\textwidth]{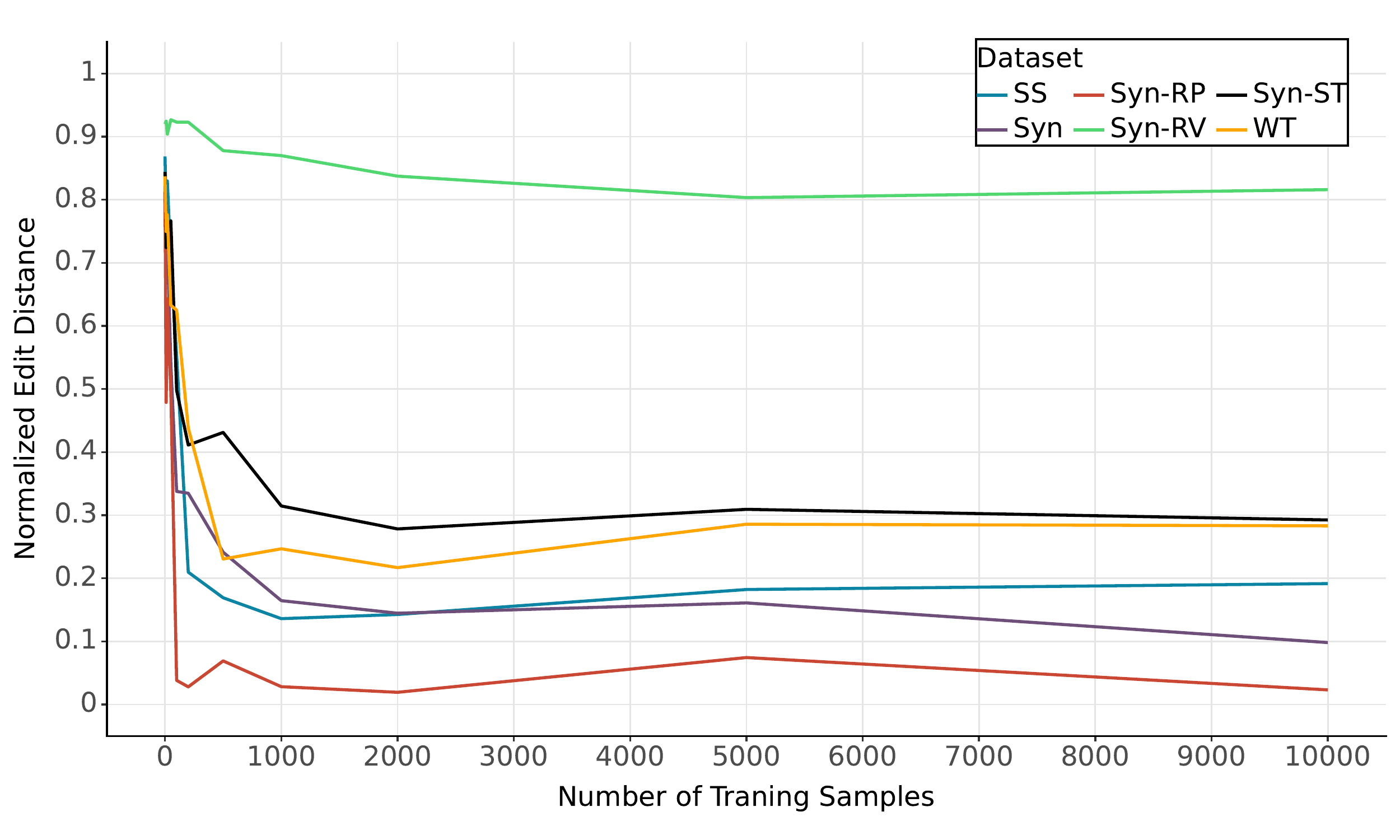}
		\caption[]{Normalized edit distance for models trained on shorter-length data} % <---
		\label{subfig:trsamp_ED_8_35}
	\end{subfigure}
	\hfill
	\begin{subfigure}[b]{0.495\textwidth}
		\centering
		\includegraphics[width=\textwidth]{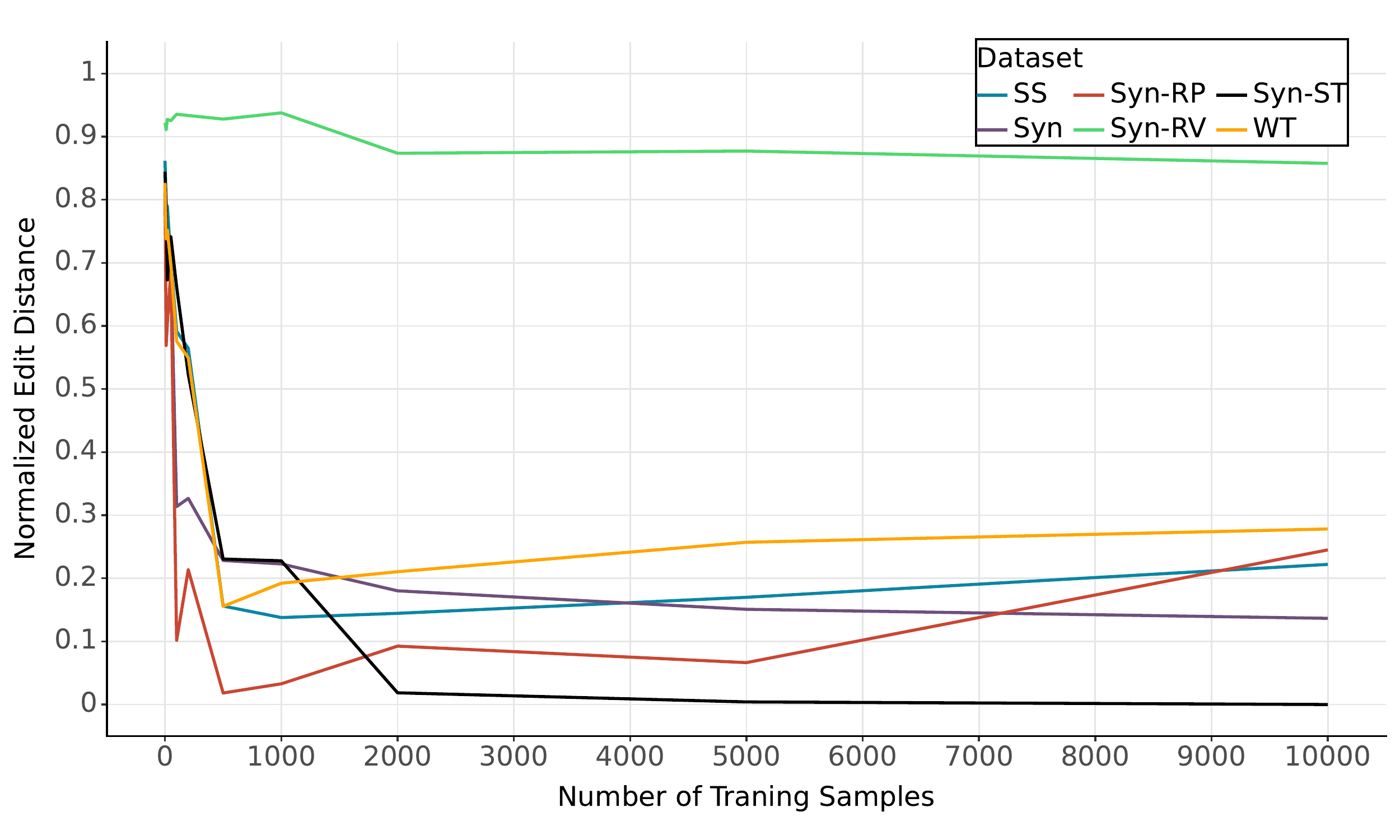}
		\caption[]{Normalized edit distance for models trained on longer-length data} % <---
		\label{subfig:trsamp_ED_5_60}
	\end{subfigure}
	\caption{Performance of the model varying the number of training data samples} % <---
	\label{fig:trsamp}
\end{figure*}

As shown in the top left panel of Figure~\ref{fig:trsamp}, the F1-Score of the model is typically less than 0.5 when no fine-tuning is performed. Also on all datasets, over 80\% of characters are predicted incorrectly (i.e. ANED $>$ 0.8) when the model is not fine-tuned, as indicated by the 0 training samples in the figure. For example, in the \textit{Syn-ST} dataset, over 84\% of output characters are predicated incorrectly by the ByT5 model without fine-tuning. However, this error is reduced to 27\% after a proper fine-tuning of the model.
This finding suggests that, unlike GPT-3, the ByT5 model without fine-tuning struggles to perform well for unequal join. Nevertheless, our fine-tuning plays a crucial role in significantly improving performance of the model.

The general expectation for the model is to perform better when more training samples are provided and the trend for our experiments is not much different. However, some observations should be taken into account. As shown in Figure~\ref{fig:trsamp}, when the number of training samples surpasses 2,000 \footnote{This number refers to the number of transformation groupings, and it translates to 20,000 source-target examples of which 16,000 examples are used for training and the remaining 4,000 is kept as the validation set.}, the model performance does not significantly change, and it reaches its optimal performance level on our datasets. Beyond this point, a slight decrease in the performance can be observed on real-world data and synthetic datasets that contain transformations not covered in the training data. This behavior can be attributed to the bias that the model acquires from seeing more transformations of the same type, which hinders its ability to effectively use its prior knowledge of real-world data. Our extensive experiments show that even with a significantly larger training dataset, the decrease in performance is not significant. Thus the model performance will converge when 2000 or more training samples are provided.

To examine how the length of input affects the training process of the model, we conducted another experiment where we changed the length range of the training samples by randomly selecting values between 5 and 60 characters.
The right panel of Figure~\ref{fig:trsamp} shows the performance when the model is trained with sequences that are generally longer and have an extended range. 
Increasing the length range of input sample pairs does not lead to any noticeable improvement on the performance of the model. That being said, increasing the length is expected to have an impact on how the model performs on longer inputs, which is discussed next.

\subsection{Performance Varying the Input Length}

In this section, we explore (a) the model's performance under different input lengths, and (b) how the length of input data during training impacts the model's ability to handle longer input during inference time. For our experiments, we regenerated our synthetic datasets under two settings, varying the input lengths between these settings. In one setting, our model was trained on input examples with lengths randomly sampled between 8 and 35, while in the other setting, the model was trained on examples with extended lengths randomly selected between 5 to 60 characters.

During inference, on datasets deemed easy in terms of the edit distance between source and target, the performance of the model is not significantly influenced by the length of the input. Both the model trained on short input examples and the model trained on long input examples deliver the highest F1-Score and near zero edit distance across almost all lengths of input.
On the medium datasets, such as \textit{Syn-ST}, the models exhibit near-perfect performance initially, and this performance is sustained for input lengths shorter than the length of the majority of samples used in training. However, a decline in  performance begins when the input length surpasses this threshold.  Interestingly, this decline in performance is not observed  when the model is trained on longer input samples. 
On the other hand, on more challenging datasets, the performance declines even for input lengths shorter than the majority of training samples. This behavior is not unexpected for Auto-regressive models, as a single incorrect prediction can influence the prediction of subsequent characters.

The results of our experiment suggest that the extent of the decrease in performance is influenced by the training data of the model. When trained on shorter-length data, there is a significant decrease in both F1-Score and ANED as the input length increases. However, when trained on lengthier data, the decrease is relatively minimal.
Overall, such cases are not very common in real-world datasets, and our experiments demonstrate that our model can perform well under various input lengths in real-world settings. 

\subsection{Robustness to Noise}

\begin{figure}[tbp]
	\centering
	\includegraphics[width=.75\linewidth]{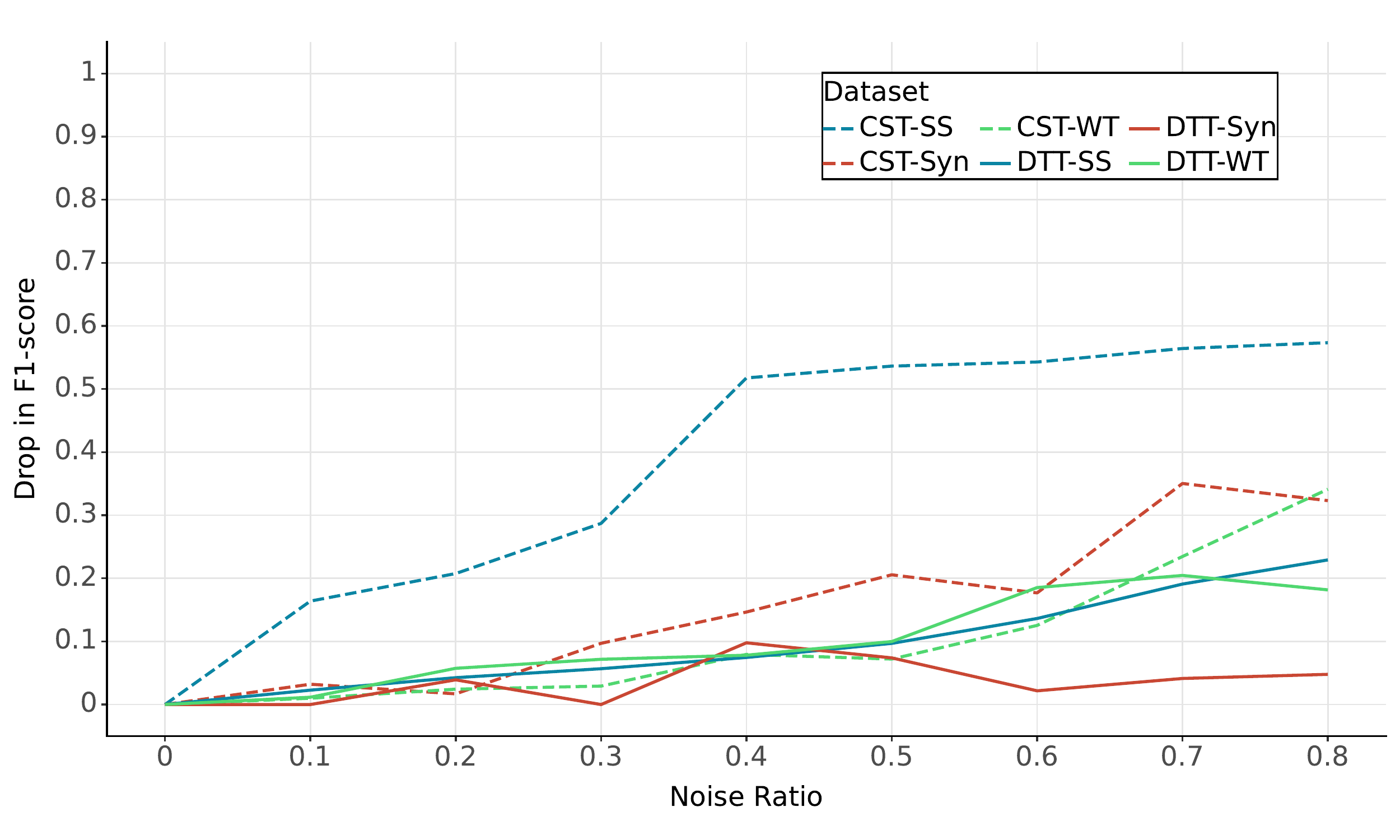}
	\caption{Performance of the model varying the ratio of noisy input samples}
	\label{fig:noise_baselines}
\end{figure}

To evaluate the impact of noisy examples on the model performance, we conducted two sets of experiments. In the first set, we introduced noise to input examples by randomly selecting  input example pairs and replacing the target with a random text. The ratio of noise was varied from 0 to 0.8 in two real-world benchmarks (\textit{WT} and \textit{SS}) and a synthetic dataset (\textit{Syn}). We utilized both DTT and CST to join the tables under noisy input samples. We compared our approach with CST as the baseline, given its recognized robustness to noise compared to other methods in the literature~\cite{icde}, including Auto-join~\cite{autojoin} and FlashFill~\cite{FlashFill,FlashFill2}.

Figure~\ref{fig:noise_baselines} demonstrates a decline in F1 score for both DTT and CST as the noise levels change. 
A lower drop in F1 score indicates greater robustness of the model to noise.
Two key observations can be drawn from this figure. Firstly, even at significantly high noise ratios of 0.7 and 0.8, the F1 score drop in our approach is less than 0.25, indicating that DTT performs very well in noisy settings. At more typical noise ratios (e.g., 0.2), the drop is less than 0.05, which is negligible in many real-world settings. 
Secondly, in the \textit{SS} and \textit{Syn} benchmarks, where the inherent noise in the data is relatively lower and textual similarity among the rows is higher, the gap between CST and DTT is noticeable even at the lowest noise ratio. This is because incorrect transformations can be generated in CST even with a single incorrect example and they will not be filtered during the join if there is high textual similarity between the pairs. On the other hand, in the \textit{WT} dataset, where the data contains inherent noise and less textual similarity between the rows, at lower noise levels, the performance of CST and DTT is almost on par, and both models can filter out invalid transformations in the join process. However, at higher noise ratios, DTT outperforms CST on the join task.

\begin{figure*}
	\begin{subfigure}[b]{0.495\textwidth}
		\includegraphics[width=\textwidth]{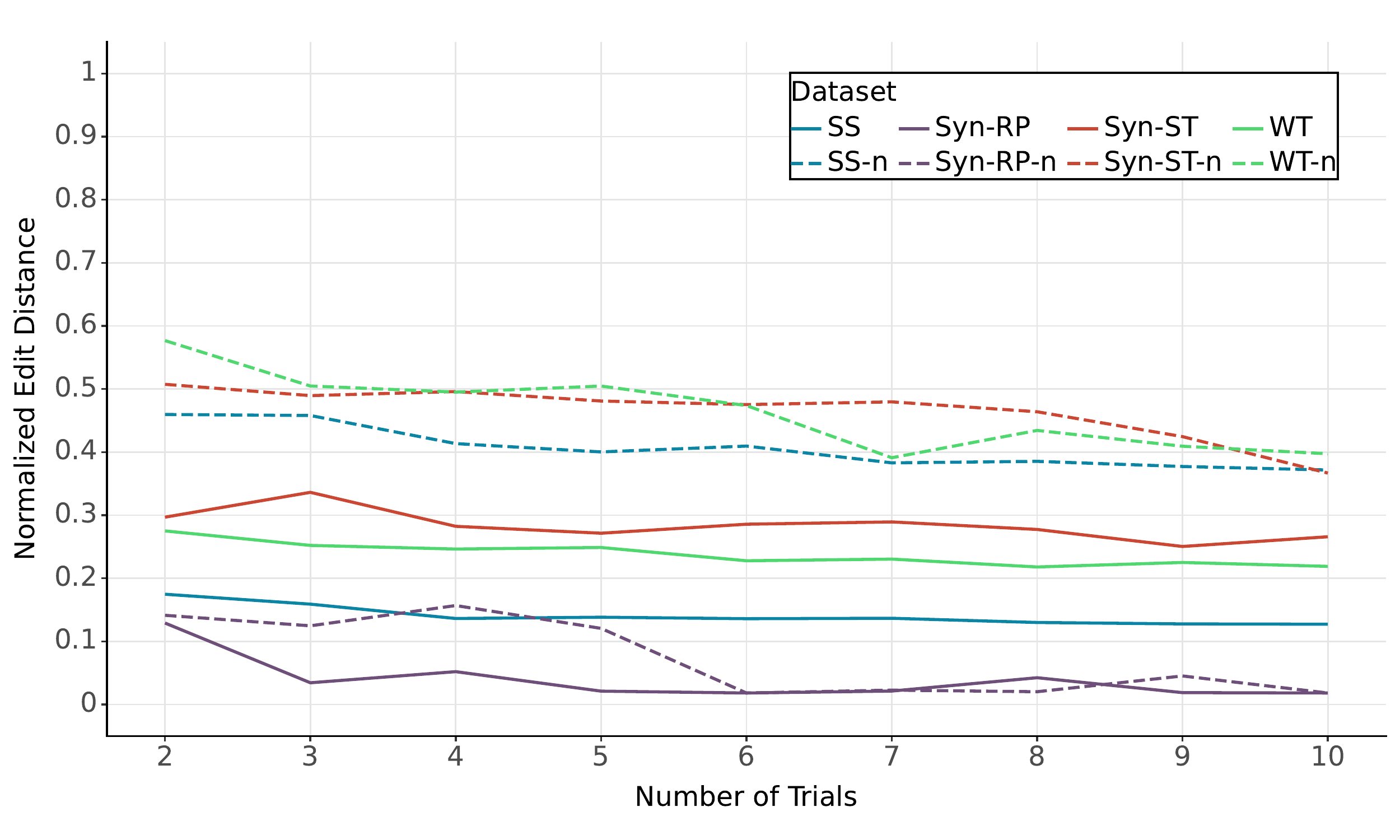}
		\caption[]{Average normalized edit distance} % <---
		\label{subfig:noise_examples_ED}
	\end{subfigure}
	\hfill
	\begin{subfigure}[b]{0.495\textwidth}
		\centering
		\includegraphics[width=\textwidth]{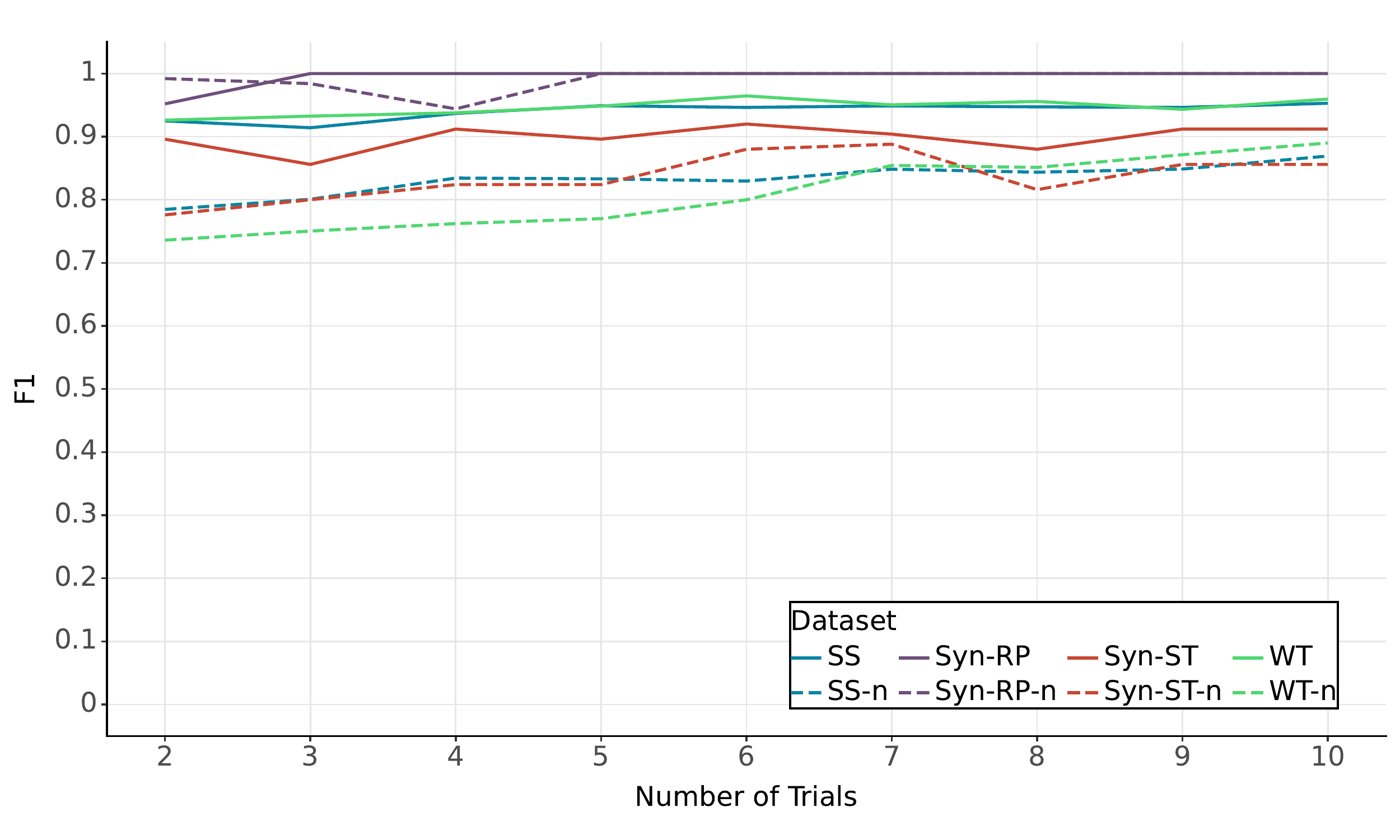}
		\caption[]{F1 score} % <---
		\label{subfig:noise_examples_F1}
	\end{subfigure}
	\caption{Impact of noise on model performance across different numbers of trials} % <---
	\label{fig:noise_example}
\end{figure*}

One of the key factors contributing to DTT's noise robustness is the decomposer and aggregator module. In the second experiment, we analyzed the effect of the number of trials in this module on the quality of transformation. Accordingly, we selected two real-world datasets (\textit{WT} and \textit{SS}) as well as two synthetic datasets (\textit{Syn-RP} and \textit{Syn-ST}), and added 60\% noise to the training examples to simulate a relatively high noise ratio. We employed the DTT framework using 2 to 10 trials for aggregation. Figure~\ref{fig:noise_example} demonstrates the performance of the model under these settings.

The left panel depicts ANED, a metric assessing the quality of the generated output (lower ANED indicates better output). On the original datasets (with no introduced noise), the performance slightly fluctuates as the number of trials increases due to inherent noise in the data and inconsistencies in model predictions. Nonetheless, the change on the noisy dataset becomes more pronounced with an increase in the number of trials, as the model has more chances to encounter correct samples and inconsistencies are better tolerated. After 5 samples with the noise level at 60\%, the performance converges and becomes closer to the scenarios with no noise. At typical noise levels of 10 to 20\%, taking 5 samples should be sufficient, which is the reason we set the number of samples to 5 in other experiments. 
Finally, the right panel displays the F1 score for the end-to-end join task under the same configuration. The trends in this chart closely mirror those of ANED. However, the changes in performance on noisy datasets are less pronounced compared to ANED, mainly because an edit-distance-based join also tolerates some noise.

\section{Conclusion and future work}
We have studied the problem of mapping tabular data from a source formatting to a desired target formatting using a set of few examples. Tables may be transformed to enable joining heterogeneous tables, filling missing values, data corrections, and other data integration tasks. To address this challenge, we proposed a framework that leverages the power of large language models. We generated the required training data and fine-tuned a character-level LLM based on ByT5 for this task. Our extensive experiments demonstrate that our model achieves impressive performance on a wide range of real-world and synthetic datasets, outperforming state-of-the-art models in the field.

Our work suggests several possible avenues for future research. One potential direction is to generalize and optimize the DTT framework for other downstream tasks such as filling missing values and error correction. Based on our experiments, in most cases the output generated by DTT is exactly equal to the expected target, making it a powerful candidate for missing value imputation.
Another direction is to explore the potentials of synthetic data generation to enhance model training for a variety of data integration tasks. There is also value in investigating the challenges and limitations of synthetic data in model training, as well as strategies for addressing those challenges.
Furthermore, given concerns around privacy, federated learning may be a preferred approach for table transformation tasks. As such, an exploration of federated learning methods for this purpose is yet another promising direction for future research. 

\section*{Acknowledgments}
This research was partially supported by the Natural Sciences and Engineering Research Council and by a grant from Servus Credit Union.

\bibliographystyle{ACM-Reference-Format}
\bibliography{refs}

\end{document}